\documentclass[pra,aps,amsfonts,amsmath,superscriptaddress,%
  twocolumn,showpacs,floatfix]{revtex4}
\usepackage{amsfonts}
\usepackage{amsmath}
\usepackage{bm}
\usepackage{epsfig}

\begin{document}

\title{\bf The Giant Monopole Resonance in Pb isotopes}.

\author{E. Khan}
\address{
{\it Institut de Physique Nucl\'eaire, Universit\'e Paris-Sud,
IN2P3-CNRS, F-91406 Orsay Cedex, France}
}


\begin{abstract}
The extraction of the nuclear incompressibility from the isoscalar
giant monopole resonance (GMR) measurements is analysed. Both pairing and
mutually enhanced magicity (MEM) effects play a role in the shift of the GMR
energy between the doubly closed shell $^{208}$Pb nucleus and other Pb
isotopes. Pairing effects are microscopically predicted whereas the MEM
effect is phenomenologically evaluated. Accurate measurements of the GMR in
open-shell Pb isotopes are called for.
\end{abstract}

\maketitle

\section{Introduction}

It has been recently shown, using the Tin isotopic chain, that pairing has
an effect on the nuclear incompressibility \cite{jli08,kha09}. In Ref.
\cite{jli08}, including pairing effects in the description of the isoscalar
giant monopole resonance (GMR) allows to explain part of the so-called Sn
softness \cite{tli07,pie07}: pairing decreases the predicted centroid of
the GMR, located at $\sim$ 16 MeV, by few hundreds of keV. In Ref.
\cite{kha09} an explicit decreasing correlation is found between the nuclei
incompressibility, obtained from the energy of the GMR, and the magnitude of
the pairing gap.

With the recent advent of accurate microscopic models in the pairing
channel, such as fully self-consistent Quasiparticle Random Phase
Approximation (QRPA) \cite{jli08,paa03}, it is now possible to predict the
GMR position with accuracy and to study small but non-negligible effects
such as pairing. Using a  microscopic approach is crucial: the GMR is
mainly built from particle-hole configurations located far from the Fermi
level, where pairing do not play a major role. However giant resonances are
known to be very collective \cite{har01} and pairing can still have a
sizable effect on the GMR properties: around 10 \% on the centroid
\cite{kha09}, which is the level of accuracy of present analysis on the
extraction of K$_\infty$ \cite{col04,vre03}. 

Experimentally, the measurement of the GMR on an isotopic chain facilitates
the study of superfluidity on the GMR properties \cite{tli07}, and the
possibility to measure the GMR in unstable nuclei emphasizes this feature
\cite{mon08}. It is therefore necessary to go towards the measurement of the
GMR on several nuclei, such as an isotopic chain. The overused method of
precise GMR measurements in a single nucleus, such as $^{208}$Pb, may not be
the relevant approach. Other nuclei have been used such as $^{90}$Zr and
$^{144}$Sm. Indeed when considering the available GMR data from which the
K$_\infty$ value has been extracted, $^{208}$Pb is stiffer than the Sn, Zr
and Sm nuclei: K$_\infty$ is about 20 MeV larger, both in non-relativistic
and in relativistic approaches \cite{col04,vre03}. The question may not be
"why Tin are so soft ?"\cite{pie07} but rather "why $^{208}$Pb is so stiff
?".  Also recent results in Cadmium isotopes confirm that open-shell nuclei
provide a value of K$_\infty$ which is lower than the one extracted from
$^{208}$Pb \cite{gar09}. 

It should be underlined that it is not possible to describe the GMR both in
Pb and in other open-shell nuclei with the same functional \cite{kha09}.
This is valid both for non relativistic and relativistic calculations. In
Ref. \cite{kha09}, it has been shown how pairing effects play a role in the
Sn isotopic chain: the energy of the GMR is increased for doubly magic
$^{132}$Sn, due to the vanishing of pairing in this nuclei. The aim of the
present work is to look for a possible similar effect in the Pb isotopic
chain. Indeed $^{204,206}$Pb nuclei are stable, but almost all the
experimental efforts in the past decades were devoted to the measurement of
the GMR in $^{208}$Pb. It would be interesting to perform accurate GMR
measurement on open-shell $^{204,206}$Pb nuclei by inelastic alpha
scattering in direct kinematics.

Even with the inclusion of pairing effects, the SLy4 functional
\cite{cha98}, which accurately describes the GMR in $^{208}$Pb, is still
overestimating the GMR in Sn isotopes (see Fig. 1 of Ref. \cite{kha09}).
Hence an additional shell effect is at work, to explain the discreapency of
the extracted K$_\infty$ values between $^{208}$Pb and Sn isotopes: the SLy4
(K$_\infty$=230 MeV) \cite{cha98} functional allows to describe the
$^{208}$Pb GMR whereas the SkM* (K$_\infty$=215 MeV) \cite{bar82} functional
is in agreement with the Sn data. 

The puzzle of the stiffness of $^{208}$Pb may come from its doubly magic
behaviour: a possible explanation is that the experimental E$_{GMR}$ data 
is especially increased in the case of doubly magic nuclei, as observed in
$^{208}$Pb compared to the GMR data available in other nuclei (such as in
the Tin isotopic chain). This difficulty to describe with a single
functional both doubly magic and other nuclei has already been observed on
the masses, namely the so-called "mutually enhancement magicity" (MEM),
described in Ref. \cite{zel83,lun03}: functionals designed to describe
masses of open-shell nuclei cannot predict the masses of doubly magic nuclei
such as $^{132}$Sn and $^{208}$Pb, which are systematically more bound that
predicted. In order to evaluate the MEM effect, it may be necessary to take
into account quadrupole correlation effects due to the flatness of the
mean-field potential for open-shell nuclei \cite{ben05}. The
incompressibility being related to the second derivative of the energy with
respect to the density, it would be useful to find a way to predict the GMR
beyond QRPA by taking into account quadrupole correlations. However such a
microscopic approach is far beyond the scope of the present work.

Therefore the MEM effect shall be phenomenologically evaluated by using a
functional which describes well the GMR in $^{208}$Pb (SLy4), and in the
case of open-shell Pb isotopes, a functional (SkM*) which describes well the
GMR in open shell nuclei, such as Sn isotopes. The aim is to predict a value
of the GMR centroid in the Pb isotopic chain which could be useful to
compare with experimental data.

\section{Constrained Hartree-Fock-Bogoliubov Calculations}

In order to consider pairing effects on the GMR, it is necessary to use a
fully microscopic method including an accurate pairing approach. We use the
constrained Hartree-Fock method, extended to the Bogoliubov pairing treatment
(CHFB) \cite{kha09,cap09}. It should be noted that the extension of the CHF
method to the CHFB case has been recently demonstrated in Ref. \cite{cap09}. The
CHF(B) method has the advantage to very precisely predict the centroid of
the GMR using the m$_{-1}$ sumrule \cite{boh79,col04b}. The whole residual
interaction (including spin-orbit and Coulomb terms) is taken into account
and this method is by construction the best to predict the GMR centroid
\cite{col04b}. Introducing the monopole operator $\hat{Q}$ as a constraint :
\begin{equation}
\hat{H}_{cons}=\hat{H}+\lambda \hat{Q}
\end{equation}
with
\begin{equation}
\hat{Q}=\sum_{i=1}^A r_i^2,
\end{equation}

the m$_{-1}$ value is obtained from the derivative of the mean value of this
operator:

\begin{equation}
m_{-1}=-\frac{1}{2}\left[\frac{\partial}{\partial\lambda}\langle
\hat{Q}\rangle\right]_{\lambda=0}
\end{equation}

The m$_1$ sumrule is extracted from the double commutator using the Thouless
theorem \cite{tho61}:
\begin{equation}
m_1=\frac{2\hbar^2}{A} \langle r^2 \rangle
\end{equation}

Finally the GMR centroid is given by E$_{GMR}$=$\sqrt{m_1/m_{-1}}$. All
details on the CHF(B) method can be found in \cite{col04,cap09,boh79}.

The present work uses the HFB approach in
coordinate space \cite{dob84} with Skyrme functionals and a zero-range
surface pairing interaction: 

\begin{equation}\label{eq:vpair}
 V_{pair}=V_0\left[1-\left(\frac{\rho(r)}{\rho_0}\right)\right]
 \delta\left({\bf r_1}-{\bf r_2}\right)
\end{equation}

This interaction is known to describe a large variety of pairing effects in
nuclei \cite{ben03}. The magnitude of the pairing interaction V$_0$ is taken
as -735 MeV.fm$^{-1}$ for SLy4 and -700 MeV.fm$^{-1}$ for SkM*: it is
adjusted so to describe the trend of the neutron pairing gap Pb isotopes.
The single quasiparticle spectrum is considered until 60 MeV. Previous CHFB
calculations in Sn isotopes are described in Ref. \cite{kha09}.

\section{Results}

\begin{figure}[htb]
\scalebox{0.35}{\includegraphics{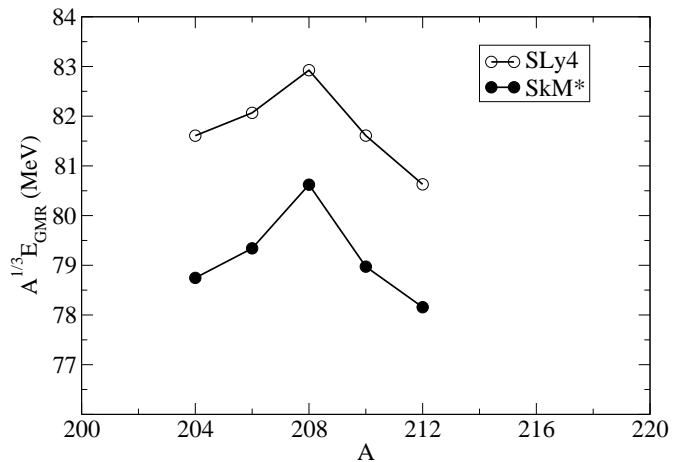}}
\caption{Excitations energies of the GMR (times A$^{1/3}$) in $^{204-212}$Pb isotopes
calculated with constrained HFB method and the SLy4 and SkM* interactions}   
\label{fig:gmr}
\end{figure}

Fig. \ref{fig:gmr} displays the GMR energy (times A$^{1/3}$ to correct for
the slow lowering of the GMR with the nuclear mass \cite{har01}) for
$^{204-212}$Pb nuclei obtained from microscopic CHFB predictions using two
functionals: SLy4 \cite{cha98} (K$_\infty$=230 MeV, which describes well the
$^{208}$Pb GMR data), and SkM* \cite{bar82} (K$_\infty$=215 MeV, which
describes well the open-shell GMR data such as Tin isotopes). As expected
the GMR energy is predicted higher in the SLy4 case than in the SkM* case.
For both interactions, the striking feature of Fig. \ref{fig:gmr} is the
increase of the GMR centroid located at the doubly magic $^{208}$Pb nucleus.
This indicates that pairing effects should be considered to describe the
behaviour of nuclear incompressibility, and that vanishing of pairing make
the nuclei stiffer to compress, confirming our previous statement on the
stiffness of $^{132}$Sn compared to open-shell Sn isotopes \cite{kha09}.
Pairing effects (CHFB calculations) decrease the centroid of the GMR as
observed in open-shell Pb isotopes, compared to $^{208}$Pb. This confirms
again the results of \cite{jli08,kha09} in the Tin data, and show that the
effect of pairing on the GMR may be universal.

We now study the hypothesis that both pairing and MEM effects are
contributing to the GMR position when comparing open-shell nuclei with
doubly magic ones. Pairing effects are treated microscopically, using the
CHFB approach, as described above. However there is no present model to take
microscopically into account the MEM effect on the GMR centroid. It should
be noted that the MEM effect is not well understood yet in the case of
nuclear masses. Nevertheless its effect on the GMR can be evaluated
phenomenologically by calculating the predicted position of the GMR centroid
with SLy4 in the $^{208}$Pb case, and with SkM* in the open-shell
$^{204,206,210,212}$Pb nuclei: SkM* allows for a good description of the GMR
in open-shell nuclei such as the Tin isotopes, where the MEM effect is at
work. As stated above, the GMR position predicted with SLy4 is in agreement with the
measurements on $^{208}$Pb, but for open-shell nuclei, a functional with
lower incompressibility should be used, as showed by the previous analysis
on Sn \cite{kha09} and Cd isotopes \cite{gar09}. We therefore use SkM* for
open-shell Pb isotopes. The aim is to provide values of the GMR in the Pb
isotopes to be compared with measurements.

\begin{figure}[htb]
\scalebox{0.35}{\includegraphics{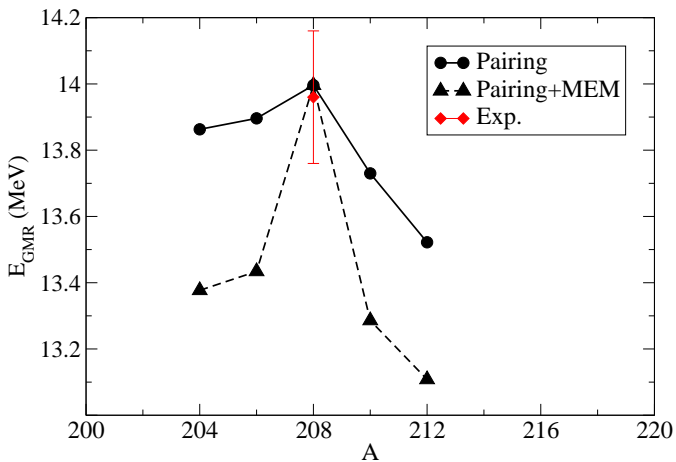}}
\caption{Excitations energies of the GMR in $^{204-212}$Pb isotopes
calculated with constrained HFB method, taking into account the MEM effect
(see text). The experimental data is taken from Ref. \cite{you04}}   
\label{fig:gmr2}
\end{figure}

Fig. \ref{fig:gmr2} displays the predicted energy of the GMR for
$^{204-212}$Pb. The solid lines corresponds to CHFB calculations using the
SLy4 functional: it displays the pairing effect on the GMR. The dashed line
corresponds to the CHFB calculations using the SLy4 functional for
$^{208}$Pb and the SkM* functional for open-shell Pb isotopes: it takes into
account both pairing and MEM effects. One observes that the main
contribution of the increase of the GMR centroid for $^{208}$Pb comes from
the MEM effect. However pairing effects still induce a decrease of the GMR
centroid for open-shell Pb isotopes. We expect measurements on $^{204}$Pb
and $^{206}$Pb to be compared with the values displayed on Fig.
\ref{fig:gmr2} in order to test both the pairing and MEM effects on the GMR
centroid.

It should be noted that the GMR has been measured in $^{206}$Pb several
decades ago \cite{har79}, providing a centroid value of 14.0 $\pm$ 0.3 MeV.
Therefore no deviation is found with respect to $^{208}$Pb. However this
measurement has been performed above 12 deg.: at such large angle, the GMR
cross section is very weak, compared to the giant quadrupole resonance (GQR)
cross section and it is delicate to extract a value of the GMR centroid
since both the GQR and the high energy background are important. Hence the
measurement was not optimal, especially when looking for typical 500 keV
effects. Therefore it would be of particular interest to measure the GMR in
$^{206}$Pb at 0 degree, allowing for a larger GMR cross section.
Furthermore, there is no GMR data for $^{204}$ Pb. The accuracy of future
GMR measurement in $^{204, 206}$Pb will be crucial. Moreover, it should be
mentioned that a somewhat lower value for the GMR centroid of $^{208}$Pb has
been found in the RCNP experiment (13.5 $\pm$ 0.2 \cite{uch03}), compared to
the Texas A\&M one \cite{you04}: there is a current debate about the reason
such variations. An accurate experiment on Pb isotopes, including $^{208}$Pb
should also help to solve this issue. 

Another concern is related to the MEM effect. If this effect is due to
quadrupole correlations coming from the flatness of the potential, as stated in
Ref. \cite{ben05}, it is expected smaller in the Pb case than in the Tin
one: $^{204,206}$ Pb are in the vicinity of a doubly magic nuclei and their
mean-field potentials are still sharp, allowing for a small MEM effect. In
the case of $^{112-124}$Sn nuclei, the potential is much flatter \cite{hil06}:
the $^{132}$Sn doubly magic nucleus is several neutrons away on the nuclear
chart. Therefore the MEM effect is expected larger in the Sn case than in
the Pb one and it may be possible that only pairing effects play
a role for the Pb isotopes, which are about 100 keV to 200 keV (solid line in
Fig. \ref{fig:gmr2}): it is crucial that future measurement are performed
within this accuracy.

In the case of the interpretation of the MEM effect from Ref. \cite{ben05}
is correct, taking it into account microscopically would necessitate to
predict the GMR with quadrupole correlations due to the flatness of the
mean-field potential in the case of open-shell nuclei. We expect it to lower
the predicted GMR value, as stated above. Therefore, the value
$K_\infty$=230 MeV extracted from the analysis of the $^{208}$Pb data should
be correct, whereas the apparent softness of open-shell nuclei like Tin, may
be artificial ($K_\infty$ typically 20 MeV lower), because the MEM effect
has not been included in all previous analyses.

\section{Conclusion}

The GMR energy of open-shell Pb nuclei is predicted significantly lower than
in the case of $^{208}$Pb. This is related to 2 effects: pairing and
mutually enhanced magicity. These two effects may also be the explanation of
the apparent softness of Sn isotopes, compared to $^{208}$Pb. The pairing
effect is evaluated microscopically whereas the MEM effect is evaluated
phenomenologically, since it is still not well characterised. A measurement
of the GMR in $^{204}$Pb and $^{206}$Pb compatible with the present
predictions would solve this softness problem. It would also mean that
K$_\infty$ should not be determined by measuring the GMR in a single nuclei
such as $^{208}$Pb, but the whole isotopic chain should be measured in order
to provide a general view on the various effects on the GMR.

Additional theoretical investigations are called for in order to predict the
GMR including the mutually enhancement magicity effect in a microscopic way.
This would necessitate an ambitious microscopic approach, trying to link
QRPA and GCM approaches. Experimentally, measurements of the GMR in unstable
nuclei should also help to investigate this issue \cite{mon08}.

\noindent{\bf Acknowledgements} The author thanks M. Fujiwara and U. Garg
for fruitful discussions about this work.

\end{document}